	\documentclass{article}

	\usepackage{spconf}
	\usepackage[utf8]{inputenc}
	\usepackage[english]{babel}
	\usepackage{times}
	\usepackage{graphicx}
	\usepackage{cite}
	\usepackage{color}
	\usepackage{tikz}\usetikzlibrary{backgrounds,calc,positioning, arrows, shapes.geometric}
\def\babs{\begin{abstract}}             \def\eabs{\end{abstract}}
\def\barr{\begin{array}}                \def\earr{\end{array}}
\def\bcc{\begin{center}}                \def\ecc{\end{center}}
\def\bdes{\begin{description}}          \def\edes{\end{description}}
\def\bdoc{\begin{document}}             \def\edoc{\end{document}}
\def\ben{\begin{enumerate}}             \def\een{\end{enumerate}}
\def\barr{\begin{array}}             \def\earr{\end{array}}
\def\beqn{\begin{eqnarray}}             \def\eeqn{\end{eqnarray}}
\def\beqnx{\begin{eqnarray*}}           \def\eeqnx{\end{eqnarray*}}
\def\bseqn{\begin{subeqnarray}}         \def\eseqn{\end{subeqnarray}}
\def\beq#1\eeq{\begin{equation}#1   \end{equation}}
\def\bal#1\eal{\begin{align}#1\end{align}}
\def\balx#1\ealx{\begin{align*}#1\end{align*}}
\def\beqx{$$}                           \def\eeqx{$$}
\def\bfig{\protect\begin{figure}}       \def\efig{\protect\end{figure}}
\def\bfigx{\protect\begin{figure*}}     \def\efigx{\protect\end{figure*}}
\def\bfigt{\protect\begin{figurette}}   \def\efigt{\protect\end{figurette}}
\def\bfl{\begin{flushleft}}             \def\efl{\end{flushleft}}
\def\bfr{\begin{flushright}}            \def\efr{\end{flushright}}
\def\bit{\begin{itemize}}               \def\eit{\end{itemize}}
\def\bmi{\begin{minipage}}              \def\emi{\end{minipage}}
\def\bfmi{\begin{fminipage}}            \def\efmi{\end{fminipage}}
\def\bpic{\begin{picture}}              \def\epic{\end{picture}}
\def\bqun{\begin{quotation}}            \def\equn{\end{quotation}}
\def\bsl{\begin{slide}}                 \def\esl{\end{slide}}
\def\btabb{\begin{tabbing}}             \def\etabb{\end{tabbing}}
\def\btabl{\begin{table}}               \def\etabl{\end{table}}
\def\btablx{\begin{table*}}             \def\etablx{\end{table*}}
\def\btabu{\begin{tabular}}             \def\etabu{\end{tabular}}
\def\btabx{\begin{tabular*}}            \def\etabx{\end{tabular*}}
\def\bbib{\begin{thebibliography}{999}} \def\ebib{\end{thebibliography}}
\def\bver{\begin{verbatim}}             \def\ever{\end{verbatim}}
\def\bca{\begin{cases}}                 \def\eca{\end{cases}}
\def\brk{\begin{remark}}                \def\erk{\end{remark}}

	\input abrege
	\input alphabet
\def\pth#1{\left(#1\right)}                \def\stdpth#1{(#1)}
\def\acc#1{\left\{#1\right\}}              \def\stdacc#1{\{#1\}}
\def\cro#1{\left[#1\right]}                \def\stdcro#1{[#1]}
\def\bars#1{\left|#1\right|}               \def\stdbars#1{|#1|}
\def\norm#1{\left\|#1\right\|}             \def\stdnorm#1{\|#1\|}
\def\scal#1{\left\langle#1\right\rangle}   \def\stdscal#1{\langle#1\rangle}
 
\def\bigpth#1{\bigl(#1\bigr)}              \def\biggpth#1{\biggl(#1\biggr)}
\def\bigacc#1{\bigl\{#1\bigr\}}            \def\biggacc#1{\biggl\{#1\biggr\}}
\def\bigcro#1{\bigl[#1\bigr]}              \def\biggcro#1{\biggl[#1\biggr]}
\def\bigbars#1{\bigl|#1\bigr|}             \def\biggbars#1{\biggl|#1\biggr|}
\def\bignorm#1{\bigl\|#1\bigr\|}           \def\biggnorm#1{\biggl\|#1\biggr\|}
\def\bigscal#1{\bigl\langle#1\bigr\rangle} \def\biggscal#1{\biggl\langle#1\biggr\rangle}

\def\Bigpth#1{\Bigl(#1\Bigr)}              \def\Biggpth#1{\Biggl(#1\Biggr)}
\def\Bigacc#1{\Bigl\{#1\Bigr\}}            \def\Biggacc#1{\Biggl\{#1\Biggr\}}
\def\Bigcro#1{\Bigl[#1\Bigr]}              \def\Biggcro#1{\Biggl[#1\Biggr]}
\def\Bigbars#1{\Bigl|#1\Bigr|}             \def\Biggbars#1{\Biggl|#1\Biggr|}
\def\Bignorm#1{\Bigl\|#1\Bigr\|}           \def\Biggnorm#1{\Biggl\|#1\Biggr\|}
\def\Bigscal#1{\Bigl\langle#1\Bigr\rangle} \def\Biggscal#1{\Biggl\langle#1\Biggr\rangle}

\def\diag{{\mathrm{diag}}}     \def\Diag#1{{\mathrm{diag}}\bigcro{#1}}
\def\tr{{\mathrm{tr}}\,}       \def\Tr#1{{\mathrm{tr}}\bigcro{#1}}
\def\rg{{\mathrm{rg}}\,}       \def\Rg#1{{\mathrm{rg}}\bigcro{#1}}
\def\esp{{\mathrm{E}}\,}       \def\Esp#1{{\mathrm{E}}\bigcro{#1}}
\def\var{{\mathrm{var}}\,}     \def\Var#1{{\mathrm{var}}\bigcro{#1}}
\def\cov{{\mathrm{cov}}\,}     \def\Cov#1{{\mathrm{cov}}\bigcro{#1}}
\def\cor{{\mathrm{cor}}\,}     \def\Cor#1{{\mathrm{cor}}\bigcro{#1}}
\def\cond{\textrm{cond}\,}     \def\Cond#1{\cond\bigpth{#1}}
\def\sinc{{\mathrm{sinc}}\,}   \def\Sinc#1{{\mathrm{sinc}}\bigcro{#1}}
\def\rang{{\mathrm{rang}}\,}   \def\Rang#1{\rang\bigcro{#1}}
\def\ker{\textrm{Ker}\,}       \def\Ker#1{\ker\bigcro{#1}}
\def\img{\textrm{Im}\,}        \def\Img#1{\img\bigcro{#1}}
\def\vect{{\mathrm{Vect}}\,}   \def\Vect#1{\vect\bigcro{#1}}
\def\sgn{{\mathrm{sgn}}}       \def\Sgn#1{\sgn\bigcro{#1}}
\def\reel{{\mathrm{Re}}}       \def\Reel#1{{\mathrm{Re}}\cro{#1}}
\def\card{{\mathrm{Card}}}     \def\Card#1{{\mathrm{Card}}\cro{#1}}
\def\sh{{\mathrm{sh}}}         \def\Sh#1{{\mathrm{sh}}\bigcro{#1}}
\def\ch{{\mathrm{ch}}}         \def\Ch#1{{\mathrm{ch}}\bigcro{#1}}
\def\th{{\mathrm{th}}}         \def\Th#1{{\mathrm{th}}\bigcro{#1}}
\def\coth{{\mathrm{coth}}}     \def\Coth#1{{\mathrm{coth}}\bigcro{#1}}
\def\logit{{\mathrm{logit}}}

\def\Cos#1{\cos\cro{#1}}
\def\Sin#1{\sin\cro{#1}}
\def\Exp#1{\exp\cro{#1}}
\def\Log#1{\log\cro{#1}}
\def\Logit#1{\logit\cro{#1}}
\def\Ln#1{\ln\cro{#1}}                  
\def\Det#1{\det\bigcro{#1}}
\def\Erf#1{\mathrm{erf}\cro{#1}} 
\def\InvErf#1{\mathrm{inverf}\cro{#1}} 
\def\Erfc#1{\mathrm{erfc}\cro{#1}}
\def\Erfcx#1{\mathrm{erfcx}\cro{#1}}
\def\Min#1{\min \cro{#1}}

\def\Prob#1{\mathrm{Pr}\cro{#1}}			
\def\ProbS#1#2{\mathrm{Pr}_{#1}\cro{#2}}	

\def\Div{{\mathrm{div}}}                                        
\def\Rotv{\overrightarrow{\mathop{{\mathrm{rot}}}}}             
\def\Gradv{\overrightarrow{\mathop{{\mathrm{grad}}}}}           

\def\IF{\text{if\:}}             \def\SI{\text{si\:}}
\def\If{\text{If\:}}             \def\Si{\text{Si\:}}
\def\AND{\text{and\:}}           \def\ET{\text{et\:}}
\def\OR{\text{or\:}}             \def\OU{\text{ou\:}}
\def\THEN{\text{then\:}}         \def\ALORS{\text{alors\:}}
                                 \def\DOU{\text{d'où\:}}
\def\WHERE{\text{where\:}}       \def\Ou{\text{où\:}}
\def\WHEN{\text{when\:}}         \def\QUAND{\text{quand\:}}
\def\FOR{\text{for\:}}           \def\POUR{\text{pour\:}}
\def\FORALL{\text{for all\:}}    \def\POURTOUT{\text{pour tout\:}}
\def\ST{\text{s.t.\:}}           \def\SC{\text{s.c.\:}}
\def\SUBJTO{\text{subject to\:}} \def\SOUSC{\text{sous contraintes\:}}
\def\OTHERWISE{\text{otherwise}} \def\SINON{\text{sinon}}
\def\WITH{\text{with\:}}         \def\AVEC{\text{avec\:}}
\def\IN{\text{in\:}}             \def\DANS{\text{dans\:}}

\def\arrayp{\renewcommand{\arraystretch}{.7}\setlength{\arraycolsep}{2pt}}
\def\tabp{\renewcommand{\arraystretch}{.7}\setlength{\tabcolsep}{2pt}}

\newsavebox{\fminibox}
\newlength{\fminilength}
\newenvironment{fminipage}[1][\linewidth]
  {\setlength{\fminilength}{#1}
   \begin{lrbox}{\fminibox}\begin{minipage}{\fminilength}}
  {\end{minipage}\end{lrbox}\noindent\fbox{\usebox{\fminibox}}}

	\def\pmu{^{-1}}

	\def\est#1{\hat{#1}} 
   \def\wh#1{\widehat{#1}}                 
   \def\wt#1{\widetilde{#1}} 

	\def\T{^\tD} 
	\def\+{^\dagger}
	\def\I{\,|\,}           
   \def\egdef{\stackrel{\Delta}{=}}

   \def\argmax{\mathop{\mathrm{arg\,max}}} 
   \def\argmin{\mathop{\mathrm{arg\,min}}} 
   \def\Argmax#1#2{\displaystyle \argmax_{#1}\left\{{#2}\right\}} 
   \def\Argmin#1#2{\displaystyle \argmin_{#1}\left\{{#2}\right\}}  %

   \def\froc#1#2{{#1/#2}}                
   \def\frOc#1#2{{#1\,\Big/#2}}                
   \def\fric#1#2{\frac1{#2}#1}
   \def\fracds#1#2{\frac{\displaystyle#1}{\displaystyle#2}}
   \def\diff#1#2{{\frac{d#1}{d#2}}}

   \def\derpar#1#2{{\frac{\partial #1}{\partial #2}}}
   \def\derpor#1#2{{\froc{\partial #1}{\partial #2}}}
   \def\parsec#1#2#3{{\frac{\partial^2 #1}{\partial #2\,\partial #3}}}
   \def\parsecd#1#2{{\frac{\partial^2 #1}{{\partial #2}^2}}}
   \def\porsec#1#2#3{{\froc{\partial^2 #1}{\partial #2\,\partial #3}}}
   \def\porsecd#1#2{{\froc{\partial^2 #1}{{\partial #2}^2}}}
   \def\pornd#1#2#3{{\froc{\partial^{#3} #1}{{\partial #2}^{#3}}}}

   \def\rond#1{\overset{\kern-0.33em~_\circ}{#1}}
   \def\rondit[#1]#2{\overset{\kern#1~_\circ}{#2}}

   \def\incirc#1{\pscirclebox[framesep=1pt]{\scriptsize#1}}
   \def\incircp#1{\pscirclebox[framesep=1pt]{\tiny#1}}
   \def\outcirc#1{\raisebox{-3pt}{\huge$×$}\hspace*{-.45cm}\text{\incirc{#1}}}

	\def\Out#1{\Red{\sout{#1}}} 
	
	\def\gX{\gamma_{\xb}}
	\def\gE{\gamma_{\eb}}
	\def\gXi{\gamma_{\xb}^{-1}}
	\def\gEi{\gamma_{\eb}^{-1}}
	\def\Sx{\rond{\Sb_\xb}}
	\def\Se{\rond{\Sb_\eb}}
	\def\Hf{\rond{\Hb}}
	\def\Sx{{\Sb_\xb}}
	\def\Se{{\Sb_\eb}}
	\def\Hf{{\Sb_\hb}}
	\def\S#1#2{\rond{\Sb_{#1}^{#2}}}
	\def\S#1#2{{\Sb_{#1}^{#2}}}
	\def\R#1#2{\Rb_{#1}^{#2}}
	\def\srp#1#2{\rond{s_{#1}^{#2}}(p)}
	\def\srp#1#2{{s_{#1}^{#2}}(p)}
	\def\ekt{\tilde{e}_{k}}
	\def\elt{\tilde{e}_{l}}
	\def\xr{\rond{\xb}}
	\def\er{\rond{\eb}}
	\def\yr{\rond{\yb}}
	\newcommand{\HSP}{\hspace{0.2cm}}
	\def\Red#1{{\color{red} #1}}
	\newtheorem{rem}{Remark}

	\title{Bayesian model selection for unsupervised image deconvolution \\ with structured Gaussian priors}
	\name{B. Harrou\'e $^{1,2}$, J.-F. Giovannelli $^{1}$ and M. Pereyra $^{2}$ }
	\address{	(1)~IMS (Univ. Bordeaux, CNRS, B-INP), Talence, France \\
				(2)~MACS, Heriot-Watt University, Edinburgh, United Kingdom}

\begin{document}


	\maketitle

\begin{abstract}
This paper considers the objective comparison of stochastic models to solve inverse problems, more specifically image restoration. Most often, model comparison is addressed in a supervised manner, that can be time-consuming and partly arbitrary. Here we adopt an unsupervised Bayesian approach and objectively compare the models based on their posterior probabilities, directly from the data without ground truth available. The probabilities depend on the marginal likelihood or ``evidence'' of the models and we resort to the Chib approach including a Gibbs sampler. We focus on the family of Gaussian models with circulant covariances and unknown hyperparameters, and compare different types of covariance matrices for the image and noise.
\end{abstract}

\section{Introduction}

Image restoration is a subject of interest in many fields: medical imaging, astronomy, physics in general, and the literature on the subject is extensive~\cite{Giovannelli15,Kaipio05}. The recurrent difficulty most often comes from the badly-scaled character, then regularization is required. Regularization can be founded on probabilistic models, such as Markov models, Gaussian models, mixtures of distributions,\dots In practice, it is necessary to know the structure of these models: neighborhood of a Markov models, type of covariance for a Gaussian, number of components of a mixture,\dots 
A common pragmatic  approach consists in giving oneself a family of models and comparing them in an empirical way. This approach has two disadvantages: it requires time to supervise the study and it is partly arbitrary. The advantage of (automatic) model selection is then obvious. 
There are many approaches~\cite{Ando10,Ding18}: posterior probabilities, information criteria (IC): AIC, BIC, Deviance IC, Predictive BIC, Generalized IC, Widely Applicable BIC,\dots 
The approach used here is optimal in the sense of the Bayesian decision~\cite{Robert07}: the model with the highest posterior probability is selected among the candidate models. These probabilities are based on an integral called \textit{evidence}, which is often difficult to compute especially in large dimensions. Here we resort to the Chib approach itself based on a Gibbs sampler. 
See also our previous papers on the subject~\cite{Vacar12,Giovannelli14c,Barbos15,Pereyra16a}. 

\pagebreak
More specifically, the restoration relies on zero-mean Gaussian models with stationary-circulant covariances and include the comparison of various types of covariance for the image and the measurement noise. See also our preliminary paper on the subject~\cite{Harroue19}.

\section{Problem statement}

Consider the Bayesian estimation of an unknown image $\xb \in \eR^P$ from a noisy and blurred observation $\yb \in \eR^P$ related to $\xb$ by a linear model $\yb=\Hb\xb+\eb$, where $\Hb\in\eR^{P\times P}$ is a known blur operator and $\eb \in \eR^P$ is additive noise. In this paper, we suppose that there are $K$ alternative models available to recover $\xb$ from $\yb$ and investigate model selection procedures to objectively compare the  models, directly from $\yb$, without ground truth available.

More precisely, we assume that $\xb$ and $\eb$ are Gaussian random vectors with mean zeros and covariance $\Rb_\xb, \Rb_\eb \in\eR^{P\times P}$. We adopt the representation $\Rb_\xb=\gXi\Cb_\xb$ and $\Rb_\eb=\gEi\Cb_\eb$, where $\Cb_\xb, \Cb_\eb \in \eR^{P\times P}$ define the covariance structure and $\gX, \gE > 0$ control the energy of $\xb$ and $\eb$. We consider that  $\gX, \gE$ are unknown and define $\gammab = [\gX,\gE]$.

We focus on the case where $\Rb_\xb$, $\Rb_\eb$ and $\Hb$ are circulant matrices diagnolizable in a discrete Fourier basis $\Fb$:
\bit
	\item $\Rb_\xb$ =  $\gXi\Fb\+ \Sx \Fb \, ,$ \HSP \HSP $\Sx$ = $ \Diag{{s_x}(p)}_{p=1\dots P} $
	\item $\Rb_\eb$ =  $\gEi\Fb\+ \Se \Fb \, ,$ \HSP \HSP $\Se$ = $ \Diag{{s_e}(p)}_{p=1\dots P} $
	\item $\Hb$ =  $\Fb\+ \Hf \Fb  \, ,$ \HSP \HSP \HSP \HSP \HSP $\,\Hf$ = $ \Diag{s_h(p)}_{p=1\dots P}$
\eit
where $\Sx$ and $\Se$ determine the power spectral density (PSD) of $\xb$ and $\eb$, up to the scale factors $\gX$ and $\gE$. 

Without loss of generality, here we consider the following four possible alternative models for $\Sx$ and $\Se$:
\beqnx
\textrm{Lorentz}	&:& 1/[(\pi \omega^2) (1 + [\nu_{h}/\omega]^2) (1 + [\nu_{v}/\omega]^2) ] \\
\textrm{Gauss}		&:& (2\pi \omega^2)^{-1} \exp [- (\nu_{h} + \nu_{v})^2 / (2\omega^2) ] \\
\textrm{Laplace}	&:& (4\omega^2)^{-1} \exp[- (|\nu_{h}| + |\nu_{v}|)/\omega ]  \\
\textrm{White}		&:& \unbb(\nu_{h},\nu_{v})
\eeqnx
where $\omega$ is a (fixed) bandwidth parameter and $(\nu_{h},\nu_{v})$ are the horizontal and vertical image frequencies. We have chosen these models because they capture a rich variety of different regularity and pixel correlation structures, see Fig.~\ref{fig:ModRxRe}. Other models could be considered too.

\bfig[htb]
\bcc
\includegraphics[width=8.5cm]{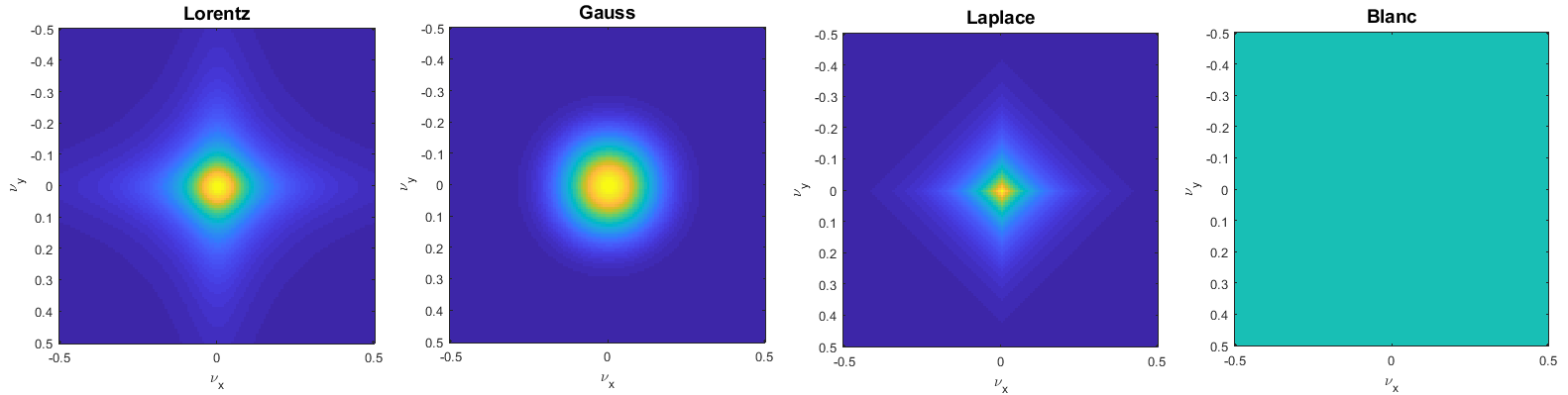}
\ecc
\vspace{-0.5cm}
\caption{Different PSD structures. From left to right: Lorentz, Gauss, Laplace and White.\label{fig:ModRxRe}}
\efig

\noindent
Moreover, we model the scale parameters $\gX,\gE$ as \aprio independent and assign them conjugate gamma density:
\beqnx
p(\gX) &\propto& \gX^{\alpha_x-1} \exp-\beta_x\gX \\
p(\gE) &\propto& \gE^{\alpha_e-1} \exp-\beta_e\gE \, ,
\eeqnx
with $\alpha_\star$ and $\beta_\star$ set to very small values to obtain vague priors.

Fig.\ref{Fig:Hierarchy} depicts graphical structure of the considered probabilistic  models.

\bfig[h]
\centering
\begin{tikzpicture}[xscale=1.25,yscale=0.8,rotate=270]

\node[draw,regular polygon,regular polygon sides=6]	{$\yb$} [grow=up]
child{node[draw,circle] {$\,\xb\,$}
	child{node[draw,circle] {$\gX$}
		child {node[draw] {$\alpha_x,\beta_x$}}
	}
}
child{node[draw,circle] {$\gE$}
	child{node[draw] {$\alpha_e,\beta_e$}}
}
;
\end{tikzpicture}
\caption{Graphical structure of the considered hierarchical model; circles represent unknown quantities.\label{Fig:Hierarchy}}
\efig

The four alternative models for $\Sx$ and $\Se$ result in $K = 16$ possible models indexed by the  variable $\Mc$ taking values in $\{1,\ldots, K\}$. Each model $\Mc$ defines a different posterior distribution for $\xb$ and will hence lead to potentially very different estimates.
The next section introduces a Bayesian approach to objectively compare the $K$ models in the absence of ground truth.

\section{Bayesian model selection by using the Chib Gibbs evidence approximation}

Following Bayesian decision theory, we compare the $K$ competing models by calculating the posterior probabilities $p(\Mc=k|\yb)$, given for any $k \in \{1,\ldots, K\}$ by
\beqn
p(\Mc=k  \I \yb )	&=& \frac{p(\yb \I \Mc=k )\, p(\Mc=k)}{p(\yb)} \nonumber\\
					&=& \frac{p(\yb \I \Mc=k )\, p(\Mc=k)}{\sum \limits_{l=1}^{K} p(\yb \I \Mc = l ) \, p(\Mc = l)}\, , \label{eq:probapost}
\eeqn
where the so-called \textsl{model evidence} or marginal likelihood
\beqx
p(\yb \I \Mc ) = \iint_{\gammab\xb} p(\yb,\xb,\gammab \I \Mc ) \, \dD\gammab \, \dD\xb \,,
\eeqx
measures the likelihood of the data given the model $\Mc$. We use the uniform prior $p(\Mc=k) = 1/K$ reflecting that all models are equally likely \aprio.

The key challenge in implementing this Bayesian decision theoretic approach is to compute model evidences. In this paper, we propose to address this difficulty by using the Chib approach~\cite{Carlin95}. More precisely, note that for all $\gammab \in \eR_{+\star}^2$
\beqn
p(\yb  \I \Mc) 	&=& \frac{ p(\yb,\gammab \I \Mc) }{p(\gammab \I \yb , \Mc)} \nonumber\\
				&=& \frac{ p(\yb \I \gammab , \Mc) \, p(\gammab \I \Mc) }{p(\gammab \I \yb,\Mc)}\label{Eq:PostProbaModel}\,,
\eeqn
and note that the numerator is tractable for the considered models. The denominator is not analytically tractable, but can be conveniently expressed as the expectation
\beqnx
p(\gammab \I \yb,  \Mc) 
	&=& \int_\xb p(\gammab,\xb \I \yb, \Mc ) \, \dD\xb \\
	&=& \int_\xb p(\gammab\I  \xb,\yb, \Mc )\,  p(\xb \I \yb,\Mc )\, \dD\xb \\
	&=& \ED_{\xb|\yb,\Mc } \cro{p(\gammab\I  \xb,\yb, \Mc )}\, ,
\eeqnx 
which can be efficiently and accurately computed by Monte Carlo integration. Precisely, we draw $G$ samples $\{\xb^{[g]}\}_{g=1}^G$ from $p(\xb \I \yb,\Mc )$ and calculate the empirical mean
\beq \label{Eq:MoyEmpirique}
\wt{p}({\gammab} \I \yb,  \Mc) 
	= \frac{1}{G} \sum_{g=1}^G p({\gammab}\I  \xb^{[g]},\yb, \Mc )\,.
\eeq 
While the above expressions are valid for all $\gammab$, they are usually evaluated at the posterior mean of $\gammab|\yb$ to reduce the variance of the empirical mean.

Markov chain Monte Carlo algorithms~\cite{Robert07,Brooks11} are a standard computation strategy to simulate the samples $\xb^{[g]}$ from $p(\xb \I \yb,\Mc )$. In particular, the Gibbs sampler is an approach of choice for the class of models considered in this paper. It iteratively constructs a Markov chain $\{\xb^{[g]},\gammab^{[g]}\}_{g=1}^G$ targeting the joint density $p(\xb,\gammab \I \yb,\Mc )$ by alternatively sampling $\gE$, $\gX$ and $\xb$ from the conditional distributions $p( \gE | \yb, \gX, \xb, \Mc)$, $p( \gX | \yb, \gE, \xb, \Mc)$ and $p( \xb | \yb, \gE, \gX, \Mc)$ evaluated at the current state of the chain.

By marginalisation through projection, the drawn samples $\{\gammab^{[g]}\}_{g=1}^G \sim p(\gammab \I \yb,\Mc )$ are used to calculate the mean $\bar{\gammab} = \sum_{g=1}^G \gammab^{[g]}/G $, followed by the computation of $\wt{p}(\bar{\gammab} \I \yb,  \Mc)$ from the samples $\{\xb^{[g]}\}_{g=1}^G \sim p(\xb \I \yb,\Mc )$ and~\eqref{Eq:MoyEmpirique}. 

Implementing this approach requires knowledge of the following five densities:

\ben
\item $p(\yb \I \gammab , \Mc=k)$ to compute the numerator~(\ref{Eq:PostProbaModel}),
\item The conditional densities defining the Gibbs sampler:
	\bit
	\item $p( \xb | \yb, \gE, \gX, \Mc)$,
	\item $p( \gE | \yb, \gX, \xb, \Mc)$,
	\item $p( \gX | \yb, \gE, \xb, \Mc)$,
	\eit
\item $p(\gammab\I  \xb,\yb, \Mc )$ to compute~(\ref{Eq:MoyEmpirique}).
\een

To derive the likelihood $p(\yb \I \gammab , \Mc=k)$ we use that $\xb$ and $\eb$ are zero-mean Gaussian vectors. Accordingly, $\yb|\gammab$ is also a Gaussian vectors with mean zero and covariance matrix
\beqx
\R{y}{k} = \Hb \R{x}{i} \Hb\+ + \R{e}{j}\, .
\eeqx
Because $\R{x}{i}$, $\R{e}{j}$ and $\Hb$ are circulant, $\R{y}{k}$ is also circulant
\beqx
	\R{y}{k} = \Fb\+ (\gXi \Hf \S{x}{i} \Hf\+ + \gEi \S{e}{j}) \Fb ~=~ \Fb \S{y}{k} \Fb\+ \,,
\eeqx
where $\S{y}{k}=\Diag{\srp{y}{k}}_{p = 1,...,P}$ is the PSD of the data and
\beqx
\srp{y}{k} = \gXi \, |s_h(p)|^2 \, \srp{x}{i} + \gEi \,  \srp{e}{j} 
\eeqx
is the variance associated to the $p$-th frequency. Moreover,
\beqx
  	\det \R{y}{k} =  \prod_{p=1}^P \srp{y}{k} 
  	~~\ET~ \yb\+ {\R{y}{k}}^{-1}\yb = \sum_{p=1}^P \frac{|\rond{y}(p)|^2}{\srp{y}{k}} \label{eq:simplif}
\eeqx
and hence the likelihood is given by
\begin{align}
&p(\yb|\gammab,\Mc) = \nonumber\\
& (2\pi)^{-P/2}\exp -\frac{1}{2}\sum_{p=1}^P \Big(\log  \srp{y}{k}  + \frac{|\rond{y}(p)|^2}{\srp{y}{k}} \Big)  \label{eq:LV}
\end{align}

The derivation of the conditional densities defining the Gibbs sampler follows from standard conjugacy results. The vector $\xb | \yb, \gE, \gX, \Mc$ is Gaussian with mean and variance
\beqnx
\mub_k	&=& \gE \, \Sigmab \, \Hb\T {\R{e}{j}}^{-1} \yb \\
\Sigmab_k	&=&  \cro{ \gE \Hb\T {\R{e}{j}}^{-1} \Hb  + {\R{x}{i}}^{-1} }\pmu\, .
\eeqnx
Similarly, the precisions  $\gE | \yb, \xb, \Mc$ and $\gX | \yb, \xb, \Mc$ are gamma densities with parameters given by
\beqx
\bca
\alpha_{e,k} &= \alpha_e + P/2\cr
\alpha_{x,k} &= \alpha_x + P/2
\eca
~~~
\bca
\beta_{e,k}  &= \beta_e + \froc{\norm{\yb - \Hb \xb}_{\R{e}{j}}^2}{2}\cr
\beta_{x,k}  &= \beta_x + \froc{\norm{\xb}_{\R{x}{i}}^2}{2} 
\eca
\eeqx

Lastly, using the fact that $\gE$ and $\gX$ are conditionally independent given $\yb, \xb, \Mc$, we obtain 
\beqx
p(\gammab \I \xb,\yb, \Mc ) = p( \gE | \yb, \xb, \Mc) ~ p( \gX | \yb, \xb, \Mc) \,.
\eeqx

Note that all the required matrix and vector products are efficiently computed by using Fourier basis representations. Similarly, one can efficiently simulate form $p( \xb | \yb, \gE, \gX, \Mc)$ by leveraging the fact that $\Sigmab_k$ is diagonal on a Fourier basis. 

\section{Numerical experiments}

We now present an experiment with synthetic data designed to demonstrate the feasibility of the proposed Bayesian model selection approach in an image processing setting.

For each of the $K = 16$  models (we have 4 alternative models for $\Sx$ and $\Se$), we generated $50$ synthetic blurred and noisy images of size $128 \times 128$. The blur is a cardinal sine of unitary width and the true values are $\gX^\star = 6$ and $\gE^\star = 4$. 

Then, for each of the $K$ (true) models and each of the $50$ images, we have computed the $K$ probabilities $p(\Mc=k  \I \yb )$ for $k = 1,\ldots, K$ by running $K$ Gibbs samplers and using the Chib approach. We then performed model selection by posterior maximisation $\hat{k} = \argmax_{k} p(\Mc=k  \I \yb )$. The results are given in Fig.~\ref{Fig:Confusion},  which shows the percentage of the number of times that each model was selected against the truth $k^\star$.

We observe that the results are very accurate for all the considered configurations, with accuracy ranging from~$90\%$ to $100\%$ depending on the specific configuration (the configurations \textsl{White/Laplace} and \textsl{White/White} seem to be particularly easy to identify, whereas the configurations \textsl{Lorentz/Gauss} and \textsl{Lorentz/White} appear to be more difficult). The overall accuracy in this experiment is over~$98\%$. 

\bfig[!h]
\bcc
\includegraphics[width=7.5cm]{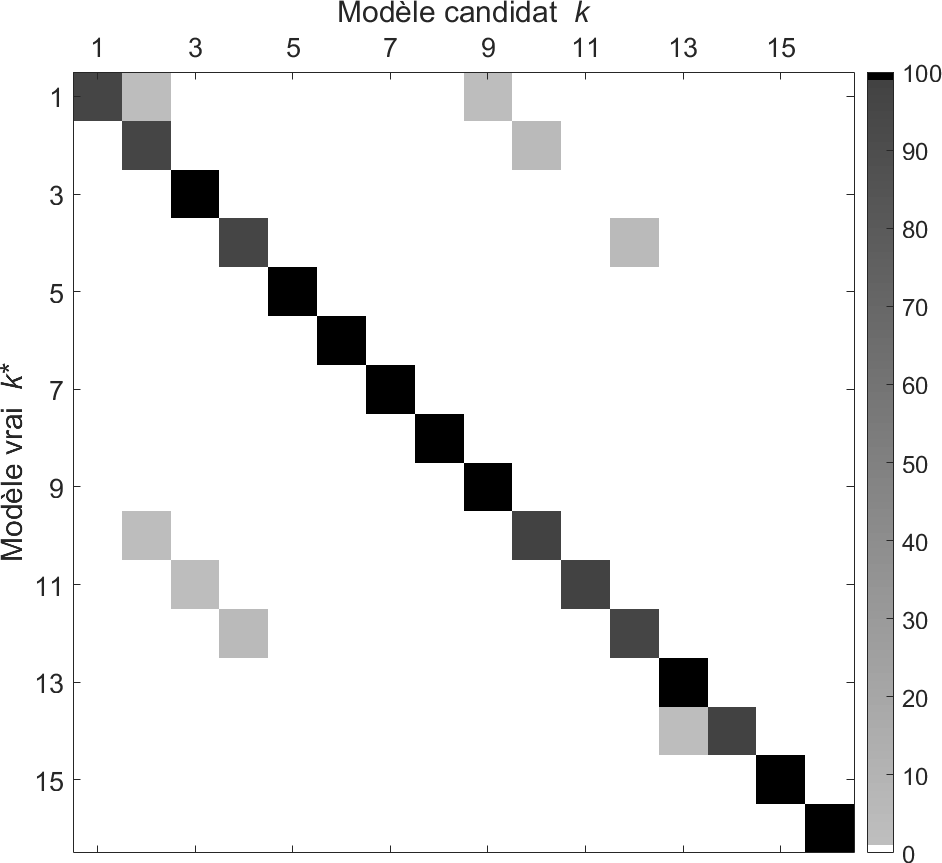}
\ecc
\caption{Confusion matrix (percentage). Candidate model (x-axis) and true model (y-axis).\label{Fig:Confusion}} 
\efig

Note that the calculation of the $K$ posterior probabilities $p(\Mc=k  \I \yb )$ for an image $\yb$ relies on $10^4$ samples, that requires nearly $15$ seconds using \texttt{MATLAB} on a standard PC because all the computations and specially the sampling under $p(\xb | \yb, \gE, \gX, \Mc)$ are performed in the Fourier domain. 

Also, the evidences are computed in logarithmic scale to avoid overflow and underflow problems. Moreover, it is important to translate the values before computing the linear scale and finally apply a factor to correct the translation. 

To produce this experiment we calculated $16\times16\times50 = 12\,800$ model evidences and never observed any convergence or numerical stability issues.

Furthermore, for illustration, Fig.~\ref{Fig:LogEvidence} shows the evolution of the approximation of the log-evidence~\eqref{Eq:PostProbaModel} based on empirical mean~\eqref{Eq:MoyEmpirique} as a function of the number $G$ of Monte Carlo samples, for one specific dataset and model configuration. For comparison, we also include the ``exact'' evidence $p(\yb \I \Mc=k)$ calculated by a computationally intensive integration of $p(\yb,\gX,\gE \I \Mc=k)$ w.r.t. $\gX,\gE$ over a fine grid.  Observe that in the order of~$10^3$ samples are required to obtain a stable approximation. It must be kept in mind that a very good precision for the log-evidence is required for a good precision on the probabilities.

\bfig[!h]
\bcc
\includegraphics[width=7.5cm]{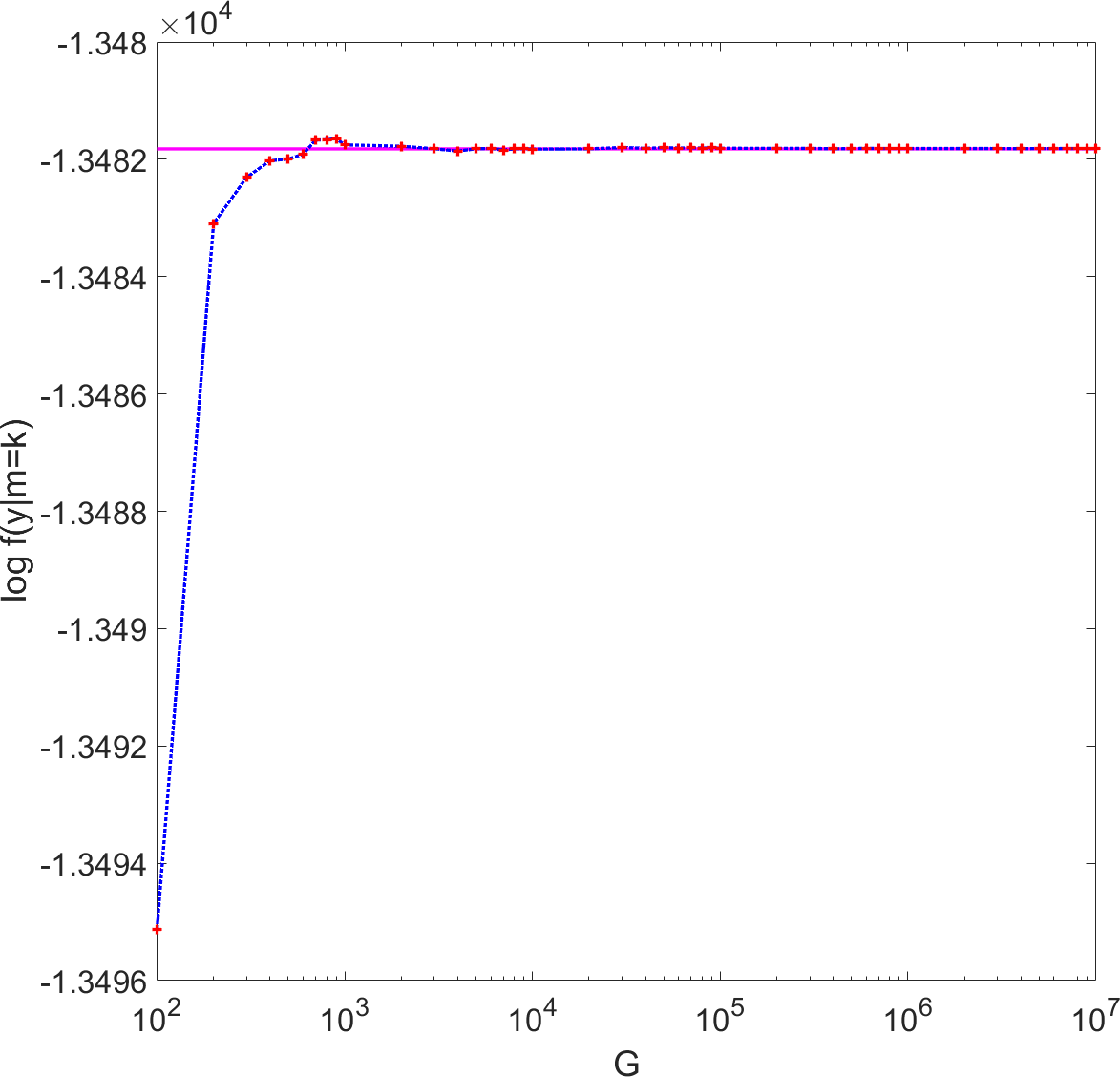}
\ecc
\vspace{-0.5cm}
\caption{Approximation of the Log-Evidence as a function of the sample number $G$. The exact value is computed by a brute force method of integration. \label{Fig:LogEvidence}}
\efig

Lastly, it is  worth noting that the Gibbs sampler also produces approximation of the marginal posteriors $p(\xb \I \yb,\Mc=k)$, $p(\gX \I \yb,\Mc=k)$ and $p(\gE \I \yb,\Mc=k)$. Fig. \ref{Fig:Chaines} shows the traces of $\gX$ and $\gE$ and the associated histograms.

\bfig[!h]
\bcc
\btabu{cc}
\includegraphics[height=2.75cm]{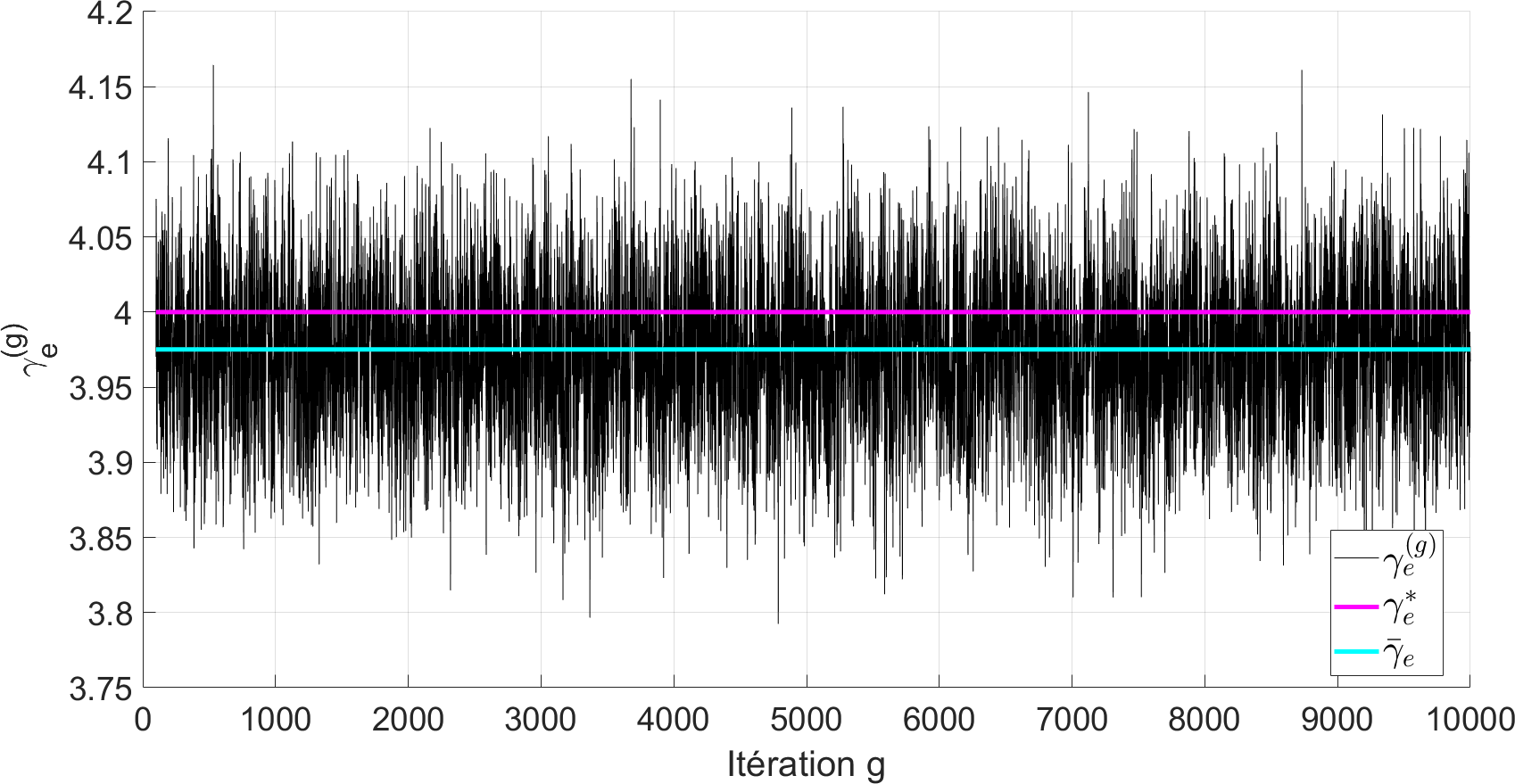} & \includegraphics[height=2.75cm]{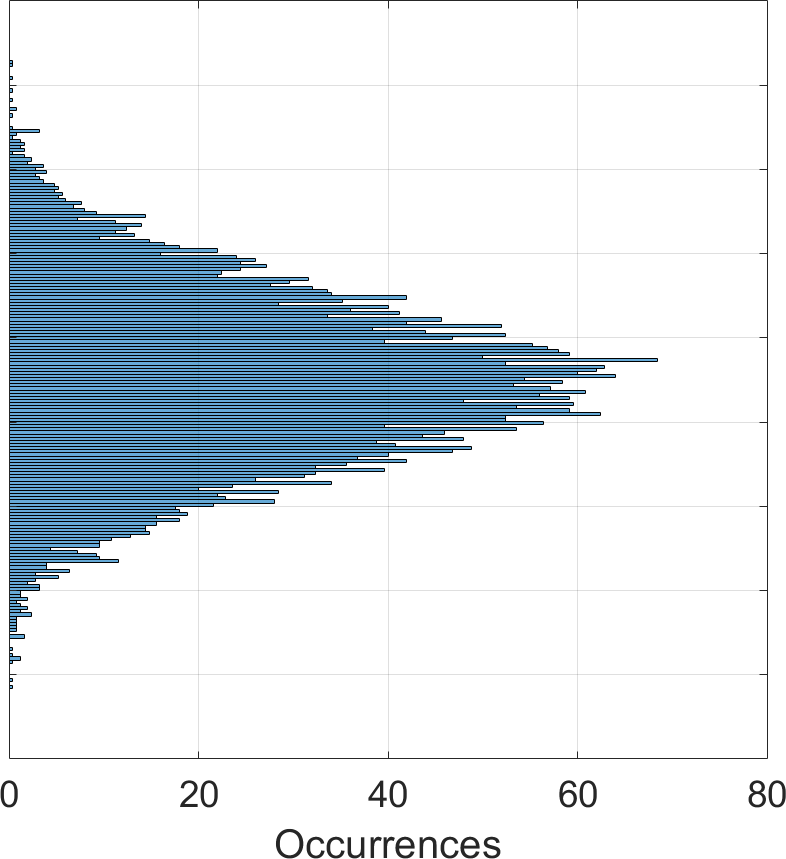} \\
\includegraphics[height=2.75cm]{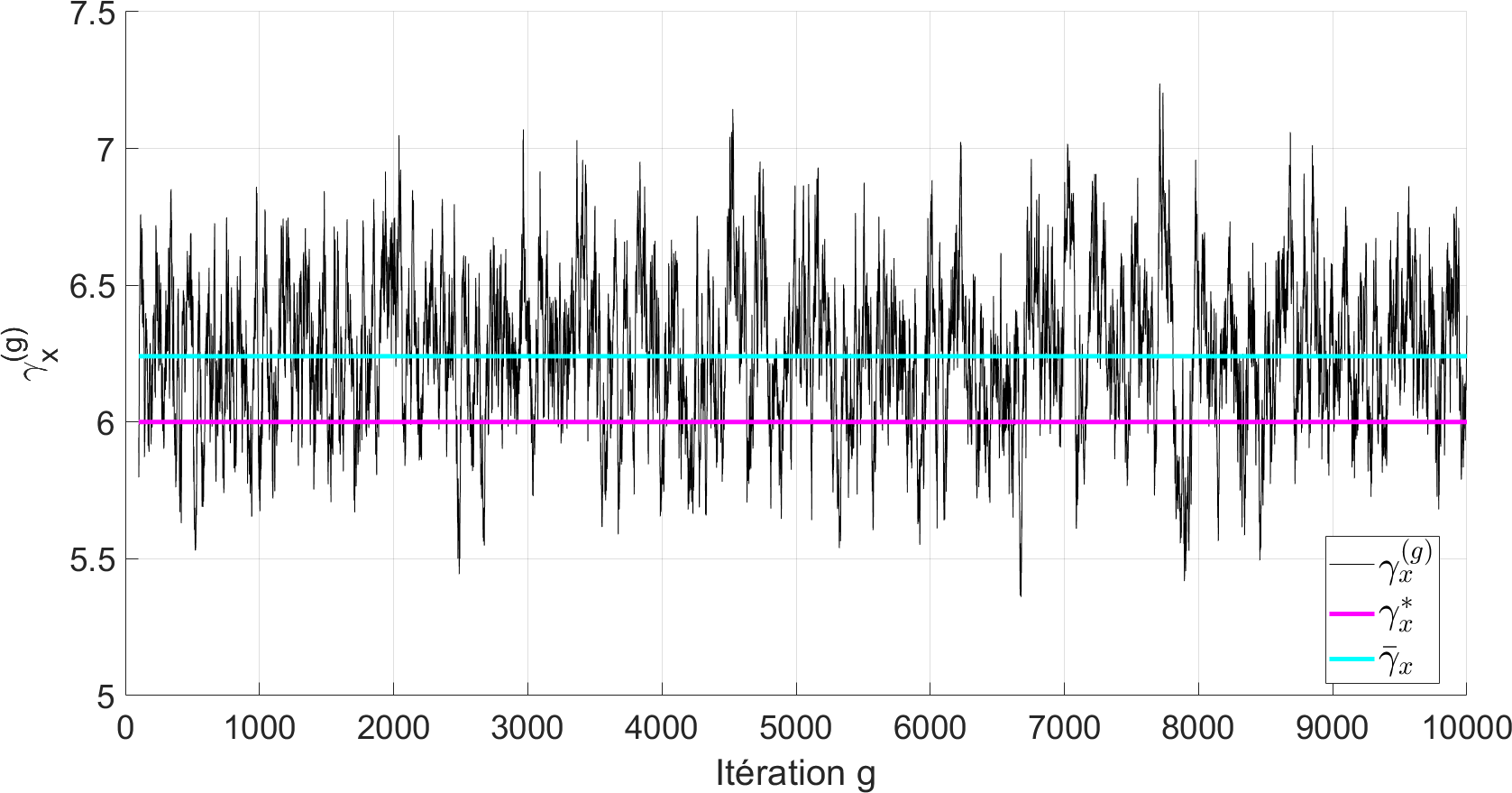} & \includegraphics[height=2.75cm]{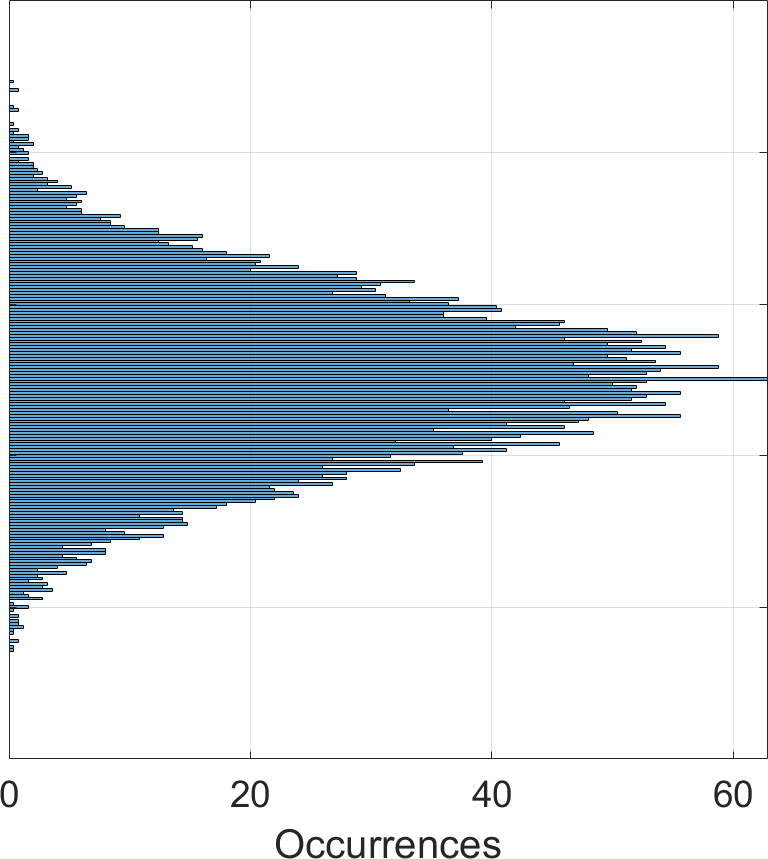} 
\etabu
\ecc
\caption{Samples of the simulated chains shown as a function the iteration index (left) and as histograms (right). Top: $\gE$ and bottom $\gX$.\label{Fig:Chaines}}
\efig

\section{Conclusion and perspectives for future work}

We have presented a contribution to the automatic selection of models for deconvolution. We have worked in the case of circulant Gaussian models which allows the marginalization of the unknown image and the easy manipulation of covariances, the major difficulty concerning the marginalization of hyperparameters. Our strategy is optimal in the sense of a Bayesian risk and is based on the choice of the most probable model. We therefore evaluate the probability of each candidate model and for this we need the evidence resulting from the marginalization of the unknown image and hyperparameters. Several options are possible and we have opted for Chib's approach and a Gibbs algorithm. We show excellent performances in terms of selection, which encourages us to extend our work. 
In an extended version of the paper, we will make a complete comparison with other existing methods: Laplace's approximation, RJMCMC, WBIC which can also be used to compute model probabilities. It will also be interesting to compare with information criteria such as AIC or BIC. 

Among the perspectives, the non-circulant Gaussian case: a direct extension will resort to~\cite{Orieux12b,Gilavert15, Marnissi18} to sample the image, the rest of the algorithm remaining unchanged. The extension to non-Gaussian cases will be based on more advanced sampling tools~\cite{Pereyra15,Pereyra16} but we will have to face a new difficulty in relation with hyperparameters and partition functions. We also intend to include new hyperparameters, \textit{e.g.} shape parameters of the image and noise DSPs, such as the width $\omega$. Naturally, processing of real data is also part of our plans.

	\small
	\bibliographystyle{IEEEtran}
	\bibliography{biben,revuedef,name,BaseBiblioAutre,BaseBiblioGPI,BaseBiblioJFG}

\end{document}